\documentclass[10pt]{article}
\usepackage{epsf,graphicx,bm}
\usepackage{wds11}

\usepackage[square,sort,comma,numbers]{natbib}

\lefthead{TSIZH ET AL.}
\righthead{Distribution of dark energy in the vicinity of compact objects}

\begin{document}

\title{Distribution of dark energy in the vicinity of compact objects}

\author{M. Tsizh, B. Novosyadlyj and Yu. Kulinich}

\affil{Astronomical Observatory of Ivan Franko National University of Lviv, Ukraine.}

\begin{abstract}

The distribution of dark energy density in the vicinity of compact static objects is analyzed. Dark energy is assumed to be in the form of a scalar field with three parameters: the background density, the equation of state parameter and the effective sound speed. Compact object is assumed to be a homogeneous spherical object of constant radius. We use the solutions of the hydrodynamical equations for dark energy in the gravitational fields of such objects for cases of static distribution of dark energy in the vicinity of star and stationary accretion onto black hole in order to analyze the possibility of constraining of the parameters of dark energy from astrophysical data. We show that dependence of density of dark energy in the vicinity of such object on the effective sound speed, background density and equation of state parameter of dark energy makes it possible to try such tests.
Here we exploit the accuracy of determination of masses of Sun and black hole in the center of Milky Way to obtain the lower limit on the effective sound speed of dark energy.
\end{abstract}

\begin{article}

\section{Introduction}
The observational data of last decade give ground to state that the Universe expands with acceleration and hence, the component with negative pressure, dubbed dark energy, must be present to provide such expansion. There are many ways to interpret theoretically this component (see, for example, \cite{Amendola,Ruiz,Novosyadlyj1}, and citation therein). One of the most developed and promising are scalar field models of dark energy.
It seems that the latest observational data prefer the phantom dark energy with EoS parameter $w<-1$ \cite{Xia,Cheng,Shafer,Hu,Novosyadlyj}. In this connection it would be useful to look for some addition possibilities to constrain the dark energy parameters, for example, analyzing the behavior of dark energy on astrophysical scales. Other motivation comes from the analysis of constraining of the effective sound speed of dark energy (\cite{sergienko} and citing therein). Current cosmological observational data do not constrain its value in the entire allowable range.

In this work we use the solutions of differential energy-momentum conservation equations for the dark energy which describe the distribution of dark energy density \cite{Tsizh} in the vicinity of spherical static objects.
The goal of this work is to study the possible manifestations of dynamical dark energy on scales of compact objects - stars, black holes etc, and try to constrain the effective sound speed of dark energy using the most accurate data on the dynamics of Solar System bodies, and data on kinematics of stars near the black hole in the center of Milky Way.

\section{Dark energy in the gravitational fields of static objects}

We analyze the behavior of dark energy in the gravitational fields of stars and black holes with the static metric in the spherical coordinates
\begin{equation}
ds^2 =  e^{\nu(r)}d\tau^2 - e^{\lambda(r)}dr^2 - r^2\left(d\theta^2+\sin^2\theta d\varphi^2\right).
\label{s2:shvartshild}
\end{equation}
We assume the scalar field model of dark energy with equation of state $p = w\rho$ ($w<-1/3$, $\rho$ is energy density) and squared effective sound speed  $c_s^2\equiv\delta p/\delta\rho=const>0$. We suppose also that at the infinite distance from the source the density and EoS parameter are constant and equal to $\rho_{\infty}$ and $w_{\infty}$ correspondingly, which we call as background values. The Lagrangian of such scalar field which is the same for cases of static world of galaxies and expanding cosmological background is presented in \cite{Tsizh,sergienko}. The equation of state (EoS) parameter for such field depends on density of
dark energy,
\begin{equation}
w=c_s^2-(c_s^2-w_{\infty})\frac{\rho_{\infty}}{\rho}.\label{w-rho}
\end{equation}
So, the pressure depends on density as $p = c_s^2\rho + w_\infty\rho_\infty$.
The components of energy-momentum tensor of such dark energy in the perfect fluid approach $T_i^k = (\rho+p)u_iu^k - \delta_i^kp$ are presented in \cite{Tsizh}. We considered two cases: the static distribution of dark energy and its stationary accretion onto black hole.

\subsection{Static distribution of dark energy density}
\begin{figure}
  \begin{center}
  \includegraphics[width=0.49\textwidth]{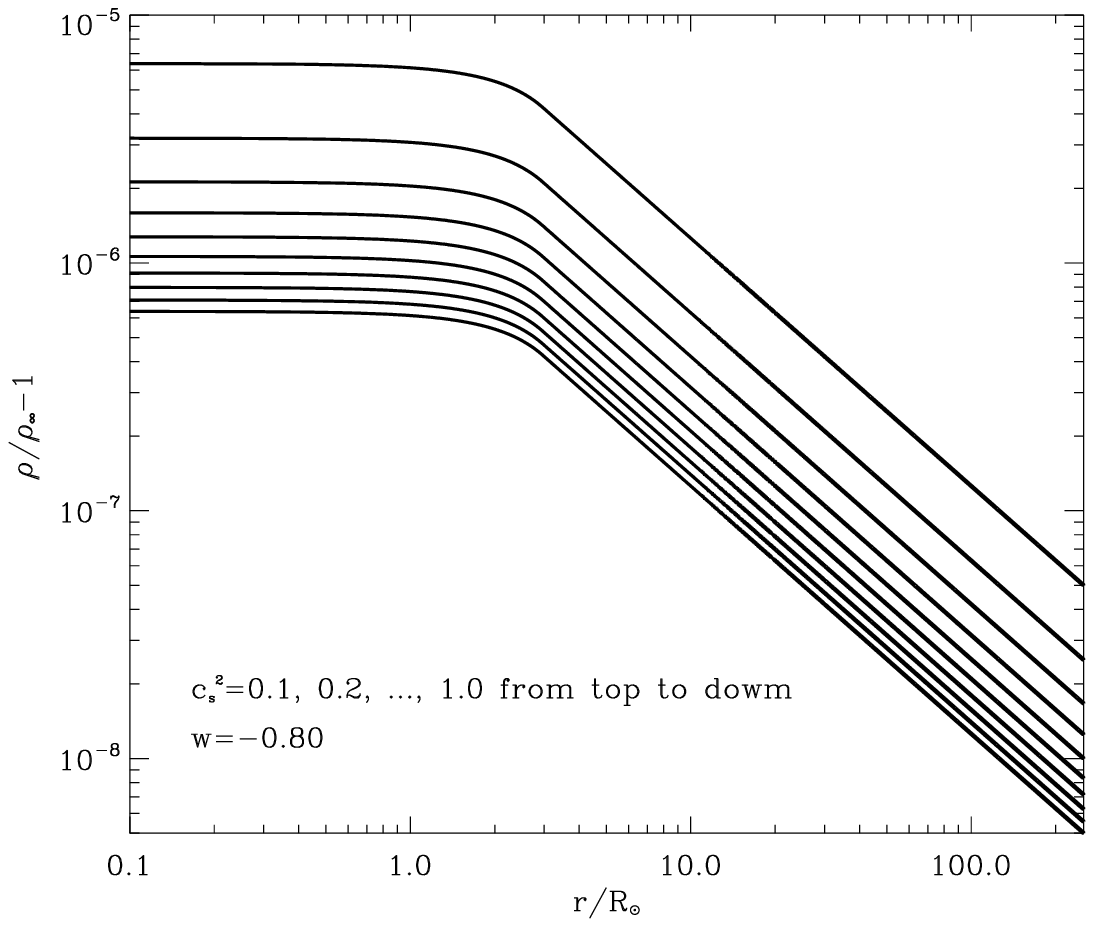}
  \includegraphics[width=0.49\textwidth]{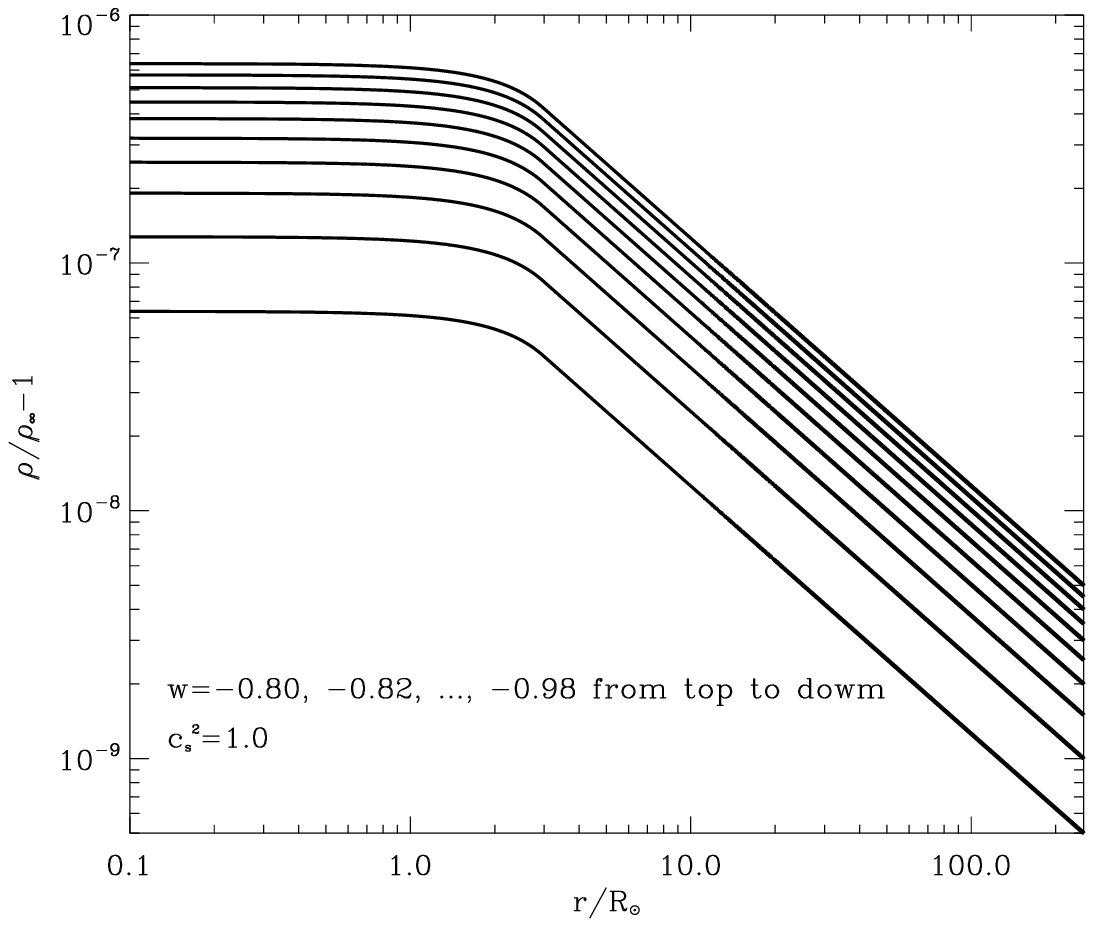}
  \end{center}
\caption{The deviation of dark energy density in the Sun-like homogeneous star and in its vicinity from the middle one in the Galaxy ($\rho(r)/\rho_{\infty}-1$) for scalar field models of dark energy with different values ​​of the effective sound speed (left) and different values ​​of EoS parameter (right). The radial coordinate is in the units of radius of star, $r/R_{\odot}$, where $R_{\odot}=696 342$ km. In the calculations we assumed also $M_{\odot}=2\cdot10^{30}$ kg.}
\label{fig3}
\end{figure}
In the case of static distribution we assume: the central compact object of mass $M$ is a homogeneous non-rotating pressureless ball of radius $R>R_g$, where $R_g=2GM/c^2$ is its gravitational radius. The Einstein equations in the metric (\ref{s2:shvartshild}) in this case give outer and intra solutions for metric functions $\nu(r)$ and $\lambda(r)$, which smoothly bind with each other at the surface:
\begin{eqnarray}
e^{\nu(r)}=e^{-\lambda(r)}=1-\frac{R_g}{r}&\quad{\rm for}\quad r \ge R& \quad{\rm and}\nonumber\\
e^{\nu(r)}=\frac{(1-\alpha)^{3/2}}{(1-\alpha r^2/R^2)^{1/2}}, \quad e^{-\lambda(r)}=1-\alpha \frac{r^2}{R^2}&\quad{\rm for}\quad r\le R&,
\label{s2:shvartshild_in}
\end{eqnarray}
where $\alpha\equiv R_g/R$ is constant for each static object. Outer solution is well known the Schwarzschild one.
The assumption about static dark energy means that its fluid velocity at any point $r$ is equal to zero, $v(r) = 0$, and dark energy density is distributed spherically symmetric and does not depend on time. In this case we have only one non-trivial equation
\begin{equation}
T^k_{1\,;k}=0: \quad \frac{dp}{dr}+\frac12(\rho+p)\frac{d\nu}{dr}=0, \label{e-c-l}
\end{equation}
which has an exact analytical solution
\begin{equation}
\rho(r)=\rho_{\infty}\left(\frac{c_s^2-w_{\infty}}{1+c_s^2}+\frac{1+w_{\infty}}{1+c_s^2}\left[e^{\nu(r)}\right]^{-\frac{1+c_s^2}{2c_s^2}}\right), \label{rho_stat3}
\end{equation}
where $e^{\nu(r)}$ is intra or outer solution (\ref{s2:shvartshild_in}) for $r\le R$ or $r\ge R$ correspondingly. The deviations of dark energy density in the Sun-like homogeneous star and in its vicinity from the background one in the Galaxy ($\rho(r)/\rho_{\infty}-1$) for scalar field models of quintessential dark energy with different values ​​of the effective sound speed and different values ​​of EoS parameter are shown in Fig. \ref{fig3}. For phantom dark energy ($w<-1$)  the deviation ($\rho(r)/\rho_{\infty}-1$) has opposite sign but the absolute values are the same for the same $|1+w|$. We note that the deviations of dark energy density in the stars and in their vicinities from the background one in the Galaxy are very small for
$c_s^2\sim 1$, but can be large for $c_s^2\rightarrow 0$. We use this property for the estimation of the lower limit for $c_s^2$
using high precision measurements in the Solar System.

\subsection{Stationary accretion onto black hole}
\begin{figure}
  \begin{center}
  \includegraphics[width=0.5\textwidth]{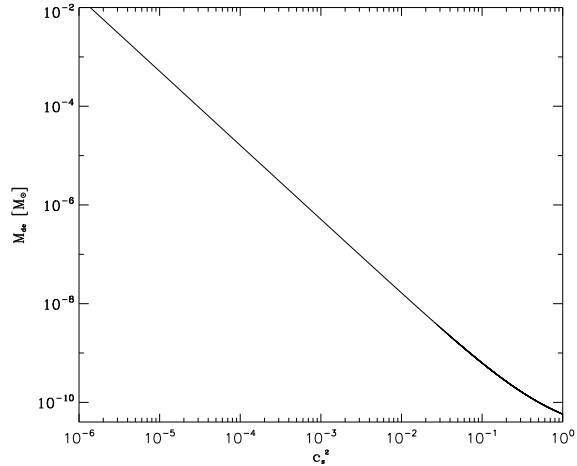}
  \end{center}
  \caption{The dependence of mass of dark energy which has flowed onto black hole during the Universe life time ($M_{de}=\dot{m}t_U$) on the effective sound speed of dark energy. The mass of black hole and dark energy model parameters are the same as in the text.}
  \label{fig4}
\end{figure}
The assumption of stationary accretion of dark energy means that it inflows in black hole vicinity but its density and flow velocity depend on radial distance and do not depend on time. The problem of accretion of gas into star in the approach of classical hydrodynamic have been solved in the works by F. Hoyle and  H. Bondi  in the middle of past century \cite{HoyLyt39,BonHoy44,Bon52}. The relativistic accretion onto black hole in the framework of General Relativity have been analyzed by Mitchel in \cite{mitchel} and many other authors reanalyzed it for different astrophysical objects and conditions. This approach have been used for analysis
of accretion of dark energy onto black hole by Babichev et al. in 2004 \cite{Babichev} (see also review \cite{Babichev2013}).
Authors concluded about decreasing of the mass of black hole due to the accretion of phantom dark energy. In \cite{Tsizh} we have re-analyzed accretion of dark energy onto black hole and show that this is fallacious conclusion despite that  the results are mathematically correct. We have obtained there the analytical solutions of hydrodynamical equations $T^i_{0\,;i}=0$ and   $u^kT^i_{k\,;i}=0$ for density and velocity of dark energy as test component in the Schwarzschild field. The analytical radial dependence of density $\rho(r)$ and 3-velocity $v(r)$ of dark energy is obtained only for some special values of $c_s^2$. We concluded that the mass of quintessential dark energy accreted onto black hole is large for dark energy model with lower value of the effective sound speed. For estimation of the amount of the dark energy in the vicinity of black hole we can use the solution for the rate of change of the inflow mass of dark energy (mass of dark energy that reached the horizon of black hole per unit
of time) which analytically depends on the dark energy parameters as well as parameters of black hole:
\begin{equation}
\dot{m}=\pi\frac{(1+3c_s^2)^{\frac{1+3c_s^2}{2c_s^2}}}{4c_s^3}(1+w_{\infty})\rho_{\infty}R_g^2.
\label{d_m}
\end{equation}
Its dependence on the value of square of effective sound speed $c_s^2$ is shown in Fig. \ref{fig4}, the dependences on the rest parameters are trivial. So, the less effective sound speed of dark energy the larger mass of dark energy that reached the horizon of black hole per finite time. This solution shows also that the rate of change of mass of the dark energy is positive in the case of quintessential dark energy and is negative in the case of the phantom one.

\section{Constraints on dark energy parameters}

It would be interesting to point out one feature of both solutions: the less sound speed $c_s$ is, the bigger changes in density near compact object and velocity of dark energy is. Knowing this, we would be able to constrain the value of effective sound speed for this model of dark energy, by comparing the limits on the uncertainties of known gravitating mass (for example, in our Solar System), supposing they are caused by accreted dark energy.

\subsection{Solar System hidden mass}
With given distribution of density near compact object one can easily find what mass of dark energy is inside sphere of certain radius. We will use our static distribution solution to find how much dark energy could it be in our Solar System. Let us integrate (\ref{rho_stat3}) to find this mass inside sphere of radius $r_1$:
\begin{eqnarray}
\label{mass}
M_{TE} (r_1) [c_s^2] = 4\pi \int_{0}^{r_1} \left(\rho + 3p\right) r^2dr =   4\pi \int_{0}^{r_1} \left(\rho(1 + 3c_s^2 ) - 3(c_s^2-w_{\infty})\rho_{\infty}\right) r^2dr =\nonumber\\
=4\pi \rho_{\infty} \Big[c_s^2 - w_{\infty})\left(\frac{1+3c_s^2}{1+c_s^2} - 3 \right)r_1^3/3 + \\
+\frac{(1+w_{\infty})(1+3c_s^2)}{1+c_s^2}\left(\int_{0}^{R_\odot}
\left[\frac{(1-\frac{R_g}{R_\odot})^{3/2}}{1-\frac{r^2 R_g}{R_\odot^3}}\right]^{-\frac{1+c_s^2}{2c_s^2}} r^2 dr + \int_{R_\odot}^{r_1} (1-\frac{R_g}{r})^{-\frac{1+c_s^2}{2c_s^2}} r^2 dr \right)\Big]\nonumber
\end{eqnarray}
where we take Solar mass and radius as mass and radius of pressureless ball. Both of integrals can not be evaluated analytically, but can be expressed through hypergeometric functions, which makes them easy to evaluate numerically. There are several attempts to find upper limit on amount of hidden mass in our Solar System: using a Pioneer effect \cite{anderson}, using Shapiro radiodelay effect of Cassini mission \cite{cassini} or using perihelion shift of objects orbiting the Sun. It looks like \cite{frere} the latter way gives the best constrains on hidden mass in Solar System. We will use the work \cite{pitjiev} on constraining amount of hidden mass based on the EPM2011 planetary ephemerides using about 677 thousand positional observations of the planets and spacecrafts. It gives, for example, not more then $5 \times 10^{-9} M_\odot$ of hidden mass inside Saturn's orbit ($r_1\sim 3\times 10^9 \,km$). Using following values of other constants
$$ R_g = 2.96 \,km,\quad R_\odot = 695.5 \cdot 10^{3} \, km $$
we find that actual constrain of speed of sound of considered dark energy model is
\begin{equation}\label{rhocs}
c_s >2\cdot10^{-4}\left[\frac{(1+w_{\infty})\rho_{\infty}}{1.2\cdot 10^{-27}}\right]^{\beta}, \qquad \beta\approx 0.17.
\end{equation}

\subsection{Supermassive black hole in the center of Milky Way}
We can try to take the advantage from the fact, that mass of the SMBH in the center of our galaxy is measured precisely enough \cite{bh}. We know, that it has mass of $(2.6\pm 0.2)\times 10^6 M_{\odot}$. We can assume that amount of dark energy near it is not larger of the uncertainty. To connect mass of dark energy with parameters we can count by multiplying rate of accretion (\ref{d_m}) by time of accretion which can be take as time of life of the Universe, $\Delta t\approx 13.6\cdot 10^9$ yr $\approx 4.2\cdot 10^{17} $ s. Hence, $c_s$ can be found from
$$\dot{m} \Delta t < 0.2 \cdot 10^6 M_{\odot} $$
This yields us
\begin{equation}
\label{rhocs2}
c_s > 4.3\cdot 10^{-6}\left[\frac{(1+w_{\infty})\rho_{\infty}}{1.2\cdot 10^{-27}} \right]^\gamma\cdot \left[\frac{ \Delta t}{4.2\cdot 10^{17}} \right]^\delta,\qquad \gamma \approx 0.51, \quad \delta\approx 0.40.
\end{equation}
All constrains are calculated at $1\sigma$ level of confidence. This constrains on the speed of sound of dark energy is important: the current uncertainties on this parameter are very large \cite{sergienko}.

\section{Conclusions}
We have solved the problem of static and stationary distributions of dark energy in the vicinity of compact objects. After obtaining such distributions we checked if dark energy can contribute significantly to gravitation in compact object systems, (for example, the Sun), and what parameters should it have to not exceed current values of errors for gravitation mass of the Sun and the SMBH in the center of Galaxy. The actual constrain of sound speed of dark energy appears to be far from the one needed in cosmology and does not rule out any model of dark energy, but at least it was shown that there is a possibility to find such restriction in principle, using observation of gravitationally bound system.

\end{article}
\end{document}